\documentclass[10pt]{article}
\usepackage{spconf,amsmath,graphicx,hyperref}
\usepackage{color}
\usepackage{cancel}
\usepackage{booktabs} 
\usepackage{subfigure}
\usepackage{verbatim}
\usepackage{siunitx}
\usepackage{url}
\usepackage{float}
\usepackage{tabularx,makecell}

\usepackage{graphicx}

\title{Pushing Bistatic Wireless Sensing toward High Accuracy at the Sub-Wavelength Scale}
\name{Wenwei Li$^{\dagger}$\qquad Jiarun Zhou$^{\dagger}$\qquad Qinxiao Quan$^{\dagger}$\qquad Fusang Zhang$^{\ddagger}$\qquad Daqing Zhang$^{\dagger*}$ \thanks{Corresponding author: Daqing Zhang. This work is supported by the National Science and Technology Major Project (Grant No. 2025ZD1302100).}}
\address{$^{\dagger}$School of Computer Science, Peking University, Beijing, China\\
         $^{\ddagger}$School of Cyber Science and Technology, Beihang University, Beijing, China\\
         $^{*}$Telecom SudParis, Institut Polytechnique de Paris, Paris, France}

\begin{document}
\small
\maketitle
\begin{abstract}
Contactless sensing using wireless communication signals has garnered significant attention due to its non-intrusive nature and ubiquitous infrastructure. Despite the promise, the inherent bistatic deployment of wireless communication introduces clock asynchronism, which leads to unknown phase offsets in channel response and hinders fine-grained sensing. State-of-the-art systems widely adopt the cross-antenna channel ratio to cancel these detrimental phase offsets. However, the channel ratio preserves sensing feature accuracy only at integer-wavelength target displacements, losing sub-wavelength fidelity. To overcome this limitation, we derive the first quantitative mapping between the distorted ratio feature and the ideal channel feature. Building on this foundation, we develop a robust framework that leverages channel response amplitude to recover the ideal channel feature from the distorted ratio. Real-world experiments across Wi-Fi and LoRa demonstrate that our method can effectively reconstruct sub-wavelength displacement details, achieving nearly an order-of-magnitude improvement in accuracy.
\end{abstract}

\section{Introduction}
Alongside the rapid proliferation of commodity wireless communication, contactless sensing leveraging ubiquitous wireless signals has attracted widespread attention over the past decade.
Compared with traditional sensing modalities that rely on dedicated sensors, wearables, or cameras, this emerging paradigm offers compelling advantages in terms of low cost, non-intrusiveness, and privacy preservation.
A wide range of signals, including Wi-Fi, 4G/5G cellular networks, and LoRa, have been explored for contactless sensing, enabling key applications such as vital sign monitoring~\cite{zeng2019farsense, zhang2021unlocking}, activity recognition~\cite{gao2022towards, chen2020robust}, and device-free tracking~\cite{li2024wifi, zhang2020exploring}.

While wireless communication systems hold great promise for sensing applications, their inherent bistatic deployment introduces a fundamental issue of clock asynchronism.
Specifically, the spatially separated transmitter and receiver typically operate on two independent oscillators, resulting in mismatches such as Carrier Frequency Offset~(CFO) and Sampling Frequency Offset~(SFO) between the two ends~\cite{kotaru2015spotfi, vasisht2016decimeter}.
These mismatches induce unknown time-varying phase offsets in the channel response measurements, which obscure the true channel phase changes induced by target motions.
Under these detrimental phase offsets, only amplitude information remains usable for sensing, significantly limiting the precision and range of potential applications~\cite{zeng2019farsense, wu2020fingerdraw}.

Prior studies have investigated various approaches to recover clean channel phase measurements for fine-grained sensing.
Early works~\cite{li2016dynamic, chen2016high, zhu2017calibrating, yu2018qgesture} employed channel correlations to estimate and reduce detrimental phase offsets, but residual offsets from estimation bias can still compromise phase accuracy.
Later efforts leveraged the cross-antenna consistency of these offsets.
Some systems~\cite{qian2017inferring, li2017indotrack, qian2018widar2, zeng2018fullbreathe} employed the cross-antenna conjugate multiplication of channel measurements to cancel the offsets, but this approach suffers from spurious motion components~\cite{li2022csi} and low SNR~\cite{wu2020fingerdraw}.
In contrast, state-of-the-art systems~\cite{zeng2019farsense, zhang2021unlocking, wu2020fingerdraw, li2022csi, chen2020robust, hu2024performance, ni2023uplink}, spanning Wi-Fi, 4G/5G, and LoRa, have widely adopted the cross-antenna channel ratio, which cancels the phase offsets while largely preserving the sensing properties of the ideal channel response.
However, although the channel ratio provides sensing features fully consistent with the ideal channel response when the target displacement is exactly an integer multiple of the wavelength, discrepancies arise at the sub-wavelength scale. 
For instance, with 2.4~GHz Wi-Fi signals of 12~cm wavelength, the channel ratio can inherently introduce sensing errors of up to 6~cm during continuous target displacements.
For signals with longer wavelengths, such as LoRa, these ratio-induced errors become even more significant.
Therefore, existing systems still face challenges in achieving high accuracy at the sub-wavelength scale.

In this paper, we present a novel framework to reconstruct sub-wavelength accuracy from the channel ratio.
To this end, we move beyond the prior qualitative understanding of the channel ratio and derive the first quantitative mapping that models the sub-wavelength feature distortion induced by the channel ratio.
{Notably, we reveal that this mapping is determined only by channel response amplitude, which is unaffected by detrimental phase offsets.}
Building on this model, we design a robust pipeline that leverages the amplitude of channel measurements to recover the ideal sensing feature from the distorted channel ratio feature.
To validate our framework, we conduct real-world experiments using both {Wi-Fi} and {LoRa} signals. 
The results show that our framework reduces the displacement estimation error of the state-of-the-art method by nearly an order of magnitude, achieving high accuracy at the sub-wavelength scale.

The rest of the paper is organized as follows. Section 2 introduces the basics of bistatic wireless sensing and the channel ratio. Section 3 presents the proposed model and framework. Section 4 reports experimental results, and Section 5 concludes the paper.
\section{Preliminary}
\subsection{Bistatic Wireless Sensing}
\begin{figure}[t]
    \centering
    \includegraphics[width=0.88\linewidth]{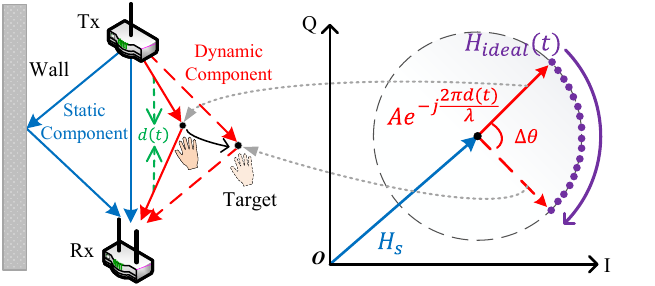}
    \vspace{-2mm}
    \caption{\small Illustration of ideal channel responses.}
    \vspace{-4mm}
    \label{fig:csi_model}
\end{figure}

In typical wireless communication environments, signals propagate from the transmitter to the receiver along multiple paths. By comparing the received signal with the known transmitted signal (\textit{e.g.}, preamble in Wi-Fi/4G/5G and dechirped symbol in LoRa), the receiver can estimate the channel response within the band of interest. 
As a superposition of components from all paths, the ideal channel response $H_{\text{ideal}}(t)$ can be expressed as
\begin{equation}
H_{\text{ideal}}(t) = \sum_{i} a_i e^{-j \frac{2\pi d_i(t)}{\lambda}} =H_s + Ae^{-j \frac{2\pi d(t)}{\lambda}},
\end{equation}
where $a_i$, $e^{-j\frac{2\pi d_i(t)}{\lambda}}$, and $d_i$ are the attenuation, phase shift, and path length of the $i$-th path signal, respectively, and $\lambda$ is the wavelength.
When a target moves within the propagation environment, the channel response can be decomposed into a static component $H_s$, resulting from the direct path and reflections from stationary objects, and a dynamic component $Ae^{-j \frac{2\pi d(t)}{\lambda}}$, contributed by the reflection from the moving target.
As target motions cause the reflection path length $d(t)$ to change, the channel response vector rotates along a circle, as shown in Figure~\ref{fig:csi_model}.
The key principle of fine-grained sensing is to extract the angle of this rotation, \textit{i.e.}, the dynamic phase change $\Delta\theta=-\frac{2\pi \Delta d}{\lambda}$, thereby quantifying the path length change $\Delta d$ and the target displacement.

In practical bistatic systems, clock asynchronism between transmitter and receiver introduces mismatches such as CFO and SFO. As a result, the actual channel response measurement $H(t)$ is impacted by an unknown time-varying phase offset $\phi_{\text{offset}}(t)$ and can be expressed as
\begin{equation}
H(t) = e^{-j\phi_{\text{offset}}(t)} \left( H_s + Ae^{-j \frac{2\pi d(t)}{\lambda}}\right).
\end{equation}
This detrimental phase offset varies significantly and obscures the phase change caused by target motions~\cite{wu2020fingerdraw, wu2021witraj}, making it challenging in practice to extract the rotation angle $\Delta\theta$ for fine-grained sensing.

\subsection{Cross-Antenna Channel Ratio}
To recover a clean channel response phase, state-of-the-art systems widely employ the cross-antenna channel ratio technique.
The underlying principle of this method is that multiple antennas on a single receiver typically share a common oscillator.
Therefore, the phase offset induced by clock asynchronism is identical across receiver antennas. 
By computing the ratio of channel measurements from two antennas, the detrimental phase offset can be perfectly canceled.
Specifically, the channel ratio $H_{\text{ratio}}(t)$ can be expressed as
{\small
\begin{equation}
\nonumber
\begin{aligned}
H_{\text{ratio}}(t) = \frac{H_1(t)}{H_2(t)} &= \frac{\cancel{e^{-j\phi_{\text{offset}}(t)}}\left( H_{s, 1} + A_1e^{-j \frac{2\pi d_1(t)}{\lambda}}\right)}{\cancel{e^{-j\phi_{\text{offset}}(t)}}\left( H_{s, 2} + A_2e^{-j \frac{2\pi d_2(t)}{\lambda}}\right)} \\
&=\frac{ H_{s, 1} + \widetilde{A_1}e^{-j \frac{2\pi d_2(t)}{\lambda}}}{ H_{s, 2} + A_2e^{-j \frac{2\pi d_2(t)}{\lambda}}}=\frac{\mathcal{A}z(t) + \mathcal{B}}{\mathcal{C}z(t) + \mathcal{D}},
\end{aligned}
\end{equation}
}
where $A_{(1, 2)}$ and $e^{-j\frac{2\pi d_{(1, 2)}(t)}{\lambda}}$ represent the amplitude and phase of the dynamic channel components at two different antennas, respectively, and $\widetilde{A_1} = A_1e^{-j\frac{2\pi (d_1-d_2)}{\lambda}}$ can be regarded as a constant over a short period~\cite{zeng2019farsense, wu2020fingerdraw}.
In this form, the channel ratio can be viewed as a Möbius transformation~\cite{needham2023visual} of $z(t)= e^{-j\frac{2\pi d_2(t)}{\lambda}}$.

A key property of the Möbius transformation is its preservation of circularity~\cite{needham2023visual}.
Since $z(t)$ traces the unit circle, the channel ratio also traces a circle.
More importantly, when $z(t)$ rotates one full cycle of $2\pi$~(\textit{i.e.}, the target path length changes by one wavelength), the channel ratio also rotates exactly one full cycle, consistent with the ideal channel response.
This property enables us to accurately estimate the path length change $\Delta d$ at the integer-wavelength scale by extracting the rotation angle of the channel ratio.


\begin{figure}[t]
    \centering
    \begin{minipage}[b]{1\linewidth}
        \centering
        \subfigure[Ideal Channel Response]{
            \centering
            \includegraphics[width = 0.45\linewidth]{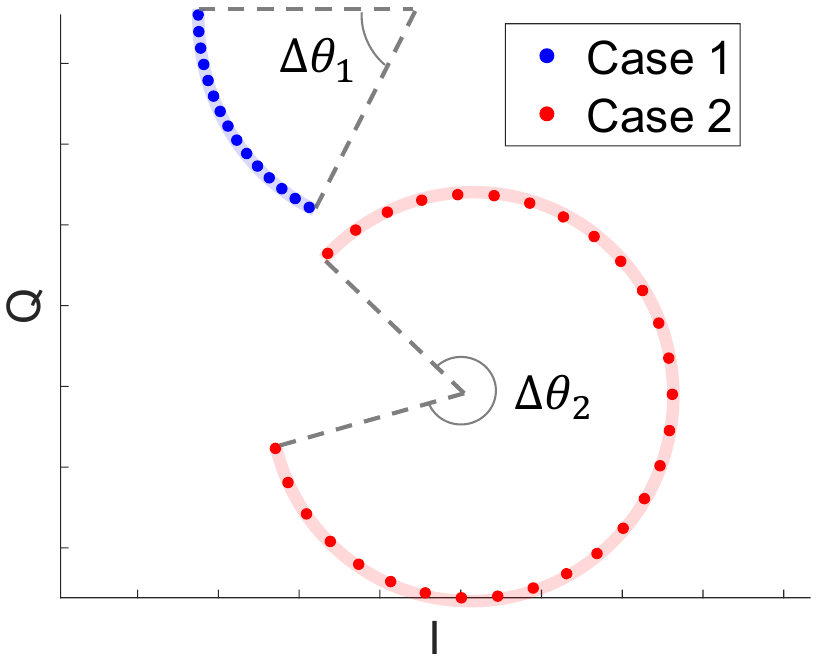}
        }
        \vspace{-1mm}
        \subfigure[Channel Ratio]{
            \centering
            \includegraphics[width = 0.45\linewidth]{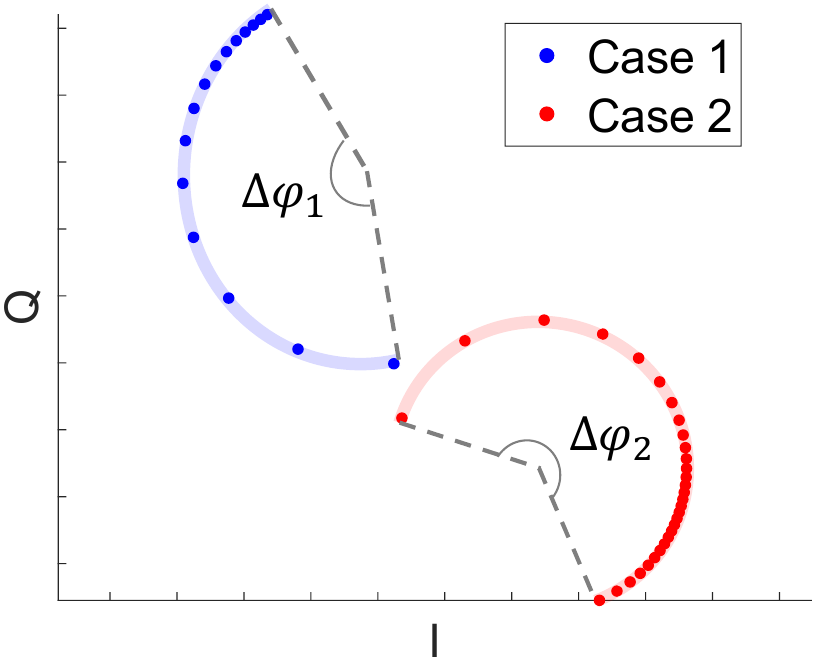}
        }
        \vspace{-1mm}
        \caption{\small The rotation feature of the channel ratio is distorted under sub-wavelength-scale variations.}
        \vspace{-3mm}
        \label{fig:csi_ratio}
    \end{minipage}
\end{figure}

Although the channel ratio enables accurate sensing at the integer-wavelength scale, it loses accuracy at the sub-wavelength scale.
As shown in Figure~\ref{fig:csi_ratio}, when the path length change $\Delta d$ is not an integer multiple of the wavelength, the rotation angle of the channel ratio may deviate significantly from that of the ideal channel response.
This deviation fundamentally limits the capability of channel-ratio–based methods to achieve sub-wavelength accuracy.
\section{Method}

\begin{figure*}[t]
    \centering
    \begin{minipage}[b]{0.315\linewidth}
        \centering
        \includegraphics[width=0.9\linewidth]{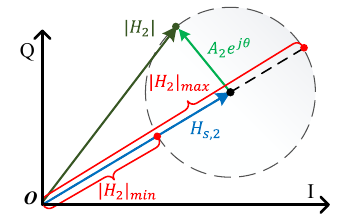}
        \vspace{-2mm}
        \caption{\small Illustration of $\left\vert H_{2}\right\vert_{\text{max}}$ and $\left\vert H_{2}\right\vert_{\text{min}}$.}
        \vspace{-3mm}
        \label{fig:equ_theta_phi}
    \end{minipage}
    \begin{minipage}[b]{0.67\linewidth}
        \centering
        \includegraphics[width=1\linewidth]{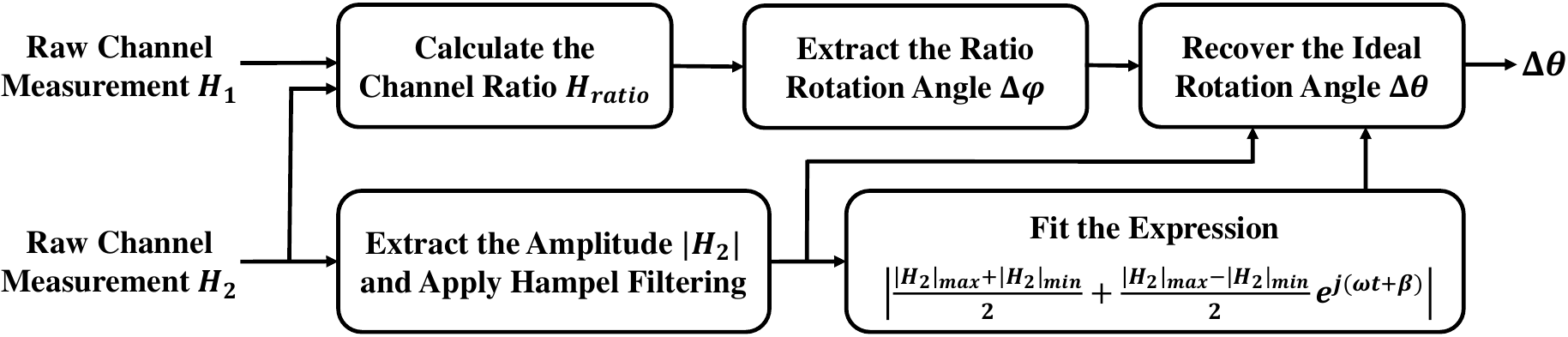}
        \vspace{-4mm}
        \caption{\small Framework Overview.}
        \vspace{-3mm}
        \label{fig:fig_4}
    \end{minipage}
\end{figure*}

In this section, we first establish a quantitative mapping between the rotation angle of the channel ratio and that of the ideal channel response. Based on this model, we propose a framework that leverages the amplitude of raw channel measurements to recover the rotation angle of the ideal channel response from the distorted channel ratio.

\subsection{Modeling the Rotation Angle of the Channel Ratio}
As the channel ratio traces a circle in the complex plane, it can be expressed as 
\begin{equation}
\label{equ:ratio}
H_{\text{ratio}} = \frac{ H_{s, 1} + \widetilde{A_1}e^{j\theta}}{ H_{s, 2} + A_2e^{j\theta}} = H_{s, \text{ratio}} + A_{\text{ratio}}e^{j\varphi}, 
\end{equation}
where $H_{s, \text{ratio}}$ and $A_{\text{ratio}}$ denote the center and the radius of the circle, respectively, $\varphi$ denote the angle of the channel ratio on the circle, and $\theta$ denotes $-\frac{2\pi d_2(t)}{\lambda}$.
Under these definitions, $\Delta\varphi$ is the rotation angle of the channel ratio, while $\Delta\theta$ is the rotation angle of the ideal channel response.

To model the rotation angle of the channel ratio, we analyze the variation of the channel ratio during the rotation. 
We assume that the variations under analysis are sufficiently small~(\textit{e.g.}, between two adjacent samples), which allows us to employ differential approximation to facilitate our analysis. 
Larger variations can be derived by accumulating such small variations.
Based on the circle-trace expression in Equation~\ref{equ:ratio}, the magnitude of the channel ratio variations can be expressed as
\begin{equation}
\label{equ:delta_ratio_1}
\left\vert \Delta H_{\text{ratio}} \right\vert = \left\vert \Delta (H_{s, \text{ratio}} + A_{\text{ratio}}e^{j\varphi}) \right\vert \approx A_{\text{ratio}} \Delta\varphi .
\end{equation}
On the other hand, as the variation of the ratio of two channel responses, its magnitude can also be expressed as
\begin{equation}
\label{equ:delta_ratio_2}
\begin{aligned}
\left\vert \Delta H_{\text{ratio}} \right\vert = \left\vert \Delta \left(\frac{ H_{s, 1} + \widetilde{A_1}e^{j\theta}}{ H_{s, 2} + A_2e^{j\theta}}\right) \right\vert &\approx \frac{\left\vert\widetilde{A_1}H_{s, 2} - A_2H_{s, 1}\right\vert}{\left\vert H_{s, 2} + A_2e^{j\theta} \right\vert^2} \Delta\theta \\
&= \frac{\left\vert\widetilde{A_1}H_{s, 2} - A_2H_{s, 1}\right\vert}{\left\vert H_2 \right\vert^2} \Delta\theta,
\end{aligned}
\end{equation}
where $\left\vert H_2 \right\vert$ represents the amplitude of the denominator channel measurement in the channel ratio.
Note that $\left\vert H_2 \right\vert$ is unaffected by detrimental phase offsets and therefore equals the amplitude of the ideal channel response.
By combining Equation~\ref{equ:delta_ratio_1} and~\ref{equ:delta_ratio_2}, we can relate $\Delta\varphi$ to $\Delta\theta$ as
\begin{equation}
\label{equ:relation_1}
\Delta\varphi = \frac{\frac{1}{A_\text{ratio}} \left\vert\widetilde{A_1}H_{s, 2} - A_2H_{s, 1}\right\vert}{\left\vert H_2 \right\vert^2} \Delta\theta = \frac{k}{\left\vert H_2 \right\vert^2} \Delta\theta,
\end{equation}
where we denote the constant term $\frac{1}{A_\text{ratio}} \left\vert\widetilde{A_1}H_{s, 2} - A_2H_{s, 1}\right\vert$ as $k$ to facilitate the subsequent analysis.

Equation~\ref{equ:relation_1} provides a basic mapping between $\Delta\varphi$ and $\Delta\theta$.
However, to reconstruct $\Delta\theta$ from $\Delta\varphi$, the constant $k$ must be expressed in terms of the accessible amplitude of channel responses.
To achieve this, we exploit the property that when the channel ratio completes a full $2\pi$ rotation, the ideal channel response also rotates exactly one full cycle.
Specifically, over a $2\pi$ rotation, the integral of the differential form on both sides of Equation~\ref{equ:relation_1} can be expressed as
\begin{equation}
\int_0^{2\pi}\mathrm{d}\varphi = \int_0^{2\pi}\frac{k}{\left\vert H_2 \right\vert^2} \mathrm{d}\theta.
\end{equation}
By substituting $\left\vert H_2 \right\vert^2 = \left\vert H_{s, 2} + A_2e^{j\theta} \right\vert^2$, $k$ can be calculated as
\begin{equation}
\begin{aligned}
k =  \frac{\int_0^{2\pi}\mathrm{d}\varphi  }{\int_0^{2\pi}\frac{1}{\left\vert H_{s, 2} + A_2e^{j\theta} \right\vert^2} \mathrm{d}\theta} &=  (\left\vert H_{s, 2} \right\vert + A_2)(\left\vert H_{s, 2}\right\vert - A_2 ) \\
&= \left\vert H_{2}\right\vert_{\text{max}}\left\vert H_{2}\right\vert_{\text{min}},
\end{aligned}
\end{equation}
where $\left\vert H_{2}\right\vert_{\text{max}}$ and $\left\vert H_{2}\right\vert_{\text{min}}$ represent the maximum and minimum value of $\left\vert H_2 \right\vert$ during the rotation of a full cycle, as illustrated in Figure~\ref{fig:equ_theta_phi}.
Finally, we derive the mapping between $\Delta\theta$ and $\Delta\varphi$ as
\begin{equation}
\label{equ:relation_2}
\Delta\theta  = \frac{\left\vert H_2 \right\vert^2}{\left\vert H_{2}\right\vert_{\text{max}}\left\vert H_{2}\right\vert_{\text{min}}} \Delta\varphi.
\end{equation}
Based on Equation~\ref{equ:relation_2}, we can reconstruct $\Delta\theta$ from $\Delta\varphi$ by extracting $\left\vert H_2 \right\vert$, $\left\vert H_{2}\right\vert_{\text{max}}$ and $\left\vert H_{2}\right\vert_{\text{min}}$ from raw channel measurements.

\subsection{Recovering the Rotation Angle of the Ideal Channel}
Figure~\ref{fig:fig_4} presents our proposed framework for recovering the rotation angle of the ideal channel response, \textit{i.e.}, $\Delta\theta$.

First, consistent with prior systems, we compute the channel ratio from the raw channel measurements of two antennas.
We then employ the state-of-the-art method proposed in~\cite{gao2022towards} to extract the rotation angle of the channel ratio, \textit{i.e.}, $\Delta\varphi$.

Concurrently, we take the raw channel measurement that serves as the denominator of the channel ratio and extract its amplitude as $\left\vert H_2 \right\vert$.
To mitigate the impact of potential impulse noise on the channel measurement amplitude ~\cite{wu2020fingerdraw}, we apply a Hampel filter to the raw $\left\vert H_2 \right\vert$.

Then, we take multiple $\left\vert H_2 \right\vert$ samples within a 0.5~s time window to extract $\left\vert H_{2}\right\vert_{\text{max}}$ and $\left\vert H_{2}\right\vert_{\text{min}}$.
A naive approach of directly taking the maximum and minimum $\left\vert H_2 \right\vert$ samples as $\left\vert H_{2}\right\vert_{\text{max}}$ and $\left\vert H_{2}\right\vert_{\text{min}}$ is not robust for two reasons: (1) these extreme values are highly susceptible to outliers and noise, and (2) during small-scale target motion, the $\left\vert H_2 \right\vert$ variation may not span its full theoretical range, meaning the true maximum and minimum may not be observed.
Therefore, we propose a scheme that leverages all samples within the window to estimate $\left\vert H_{2}\right\vert_{\text{max}}$ and $\left\vert H_{2}\right\vert_{\text{min}}$.
Specifically, since the channel response can be regarded as rotating uniformly within a short time window, the amplitude $\left\vert H_2 \right\vert$ can be modeled as
\begin{equation}
\nonumber
\left\vert \frac{\left\vert H_{2}\right\vert_{\text{max}} + \left\vert H_{2}\right\vert_{\text{min}}}{2} + \frac{\left\vert H_{2}\right\vert_{\text{max}} - \left\vert H_{2}\right\vert_{\text{min}}}{2}e^{j(\omega t+\beta)} \right\vert.
\end{equation}
We fit this expression to multiple samples within the window using a non-linear least squares algorithm, thereby obtaining robust estimates of $\left\vert H_{2}\right\vert_{\text{max}}$ and $\left\vert H_{2}\right\vert_{\text{min}}$.

Finally, we feed the extracted $\Delta\varphi$, $\left\vert H_2 \right\vert$, $\left\vert H_{2}\right\vert_{\text{max}}$ and $\left\vert H_{2}\right\vert_{\text{min}}$ into Equation~\ref{equ:relation_2} to compute $\Delta\theta=-2\pi\Delta d/\lambda$. 
We can then quantify the path length change induced by target motion with sub-wavelength accuracy.

\begin{figure*}[t]
    \centering
    \begin{minipage}[b]{0.32\linewidth}
        \centering
        \includegraphics[width=0.85\linewidth]{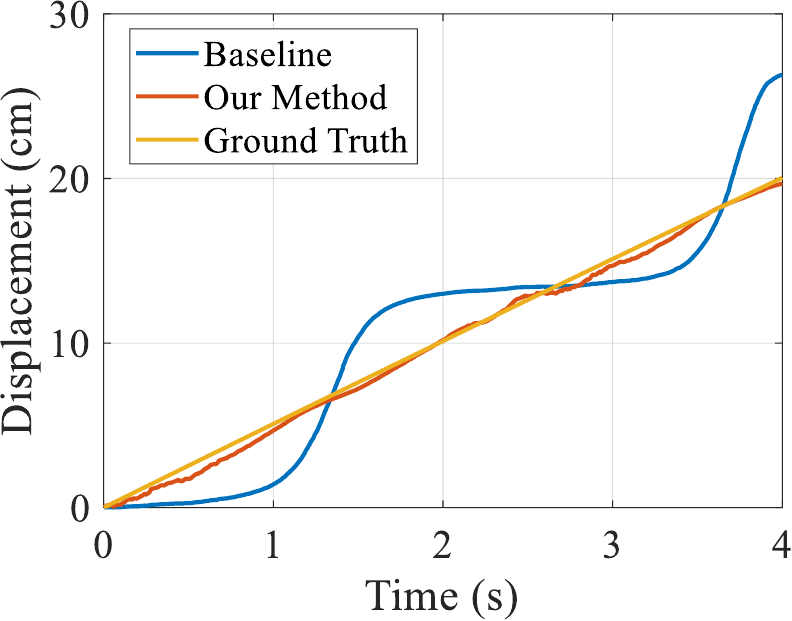}
        \vspace{-1mm}
        \caption{\small Examples of displacement estimates.}
        \vspace{-0mm}
        \label{fig:fig_5}
    \end{minipage}
    \hspace{0.01\linewidth}
    \begin{minipage}[b]{0.32\linewidth}
        \centering
        \includegraphics[width=0.85\linewidth]{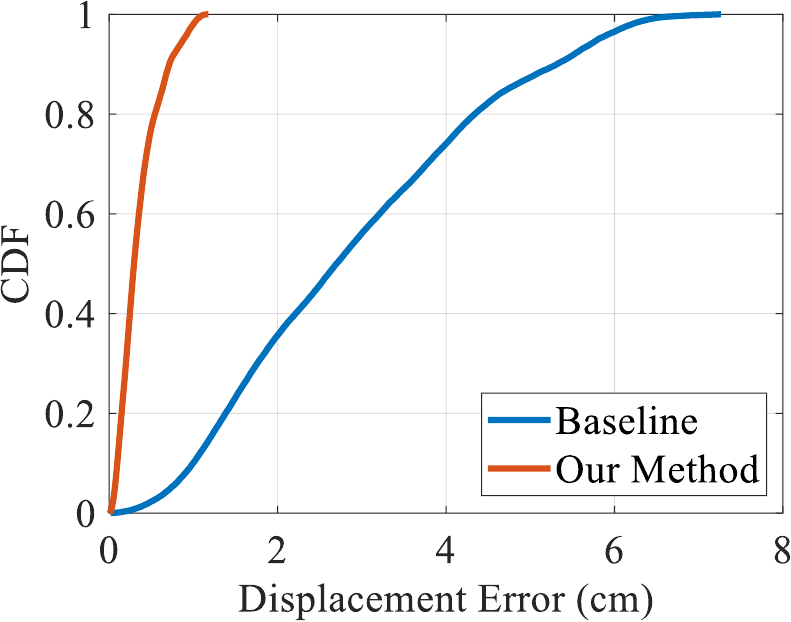}
        \vspace{-1mm}
        \caption{\small Benchmark experiment results.}
        \vspace{-0mm}
        \label{fig:fig_6}
    \end{minipage}
    \hspace{0.01\linewidth}
    \begin{minipage}[b]{0.32\linewidth}
        \centering
        \includegraphics[width=0.85\linewidth]{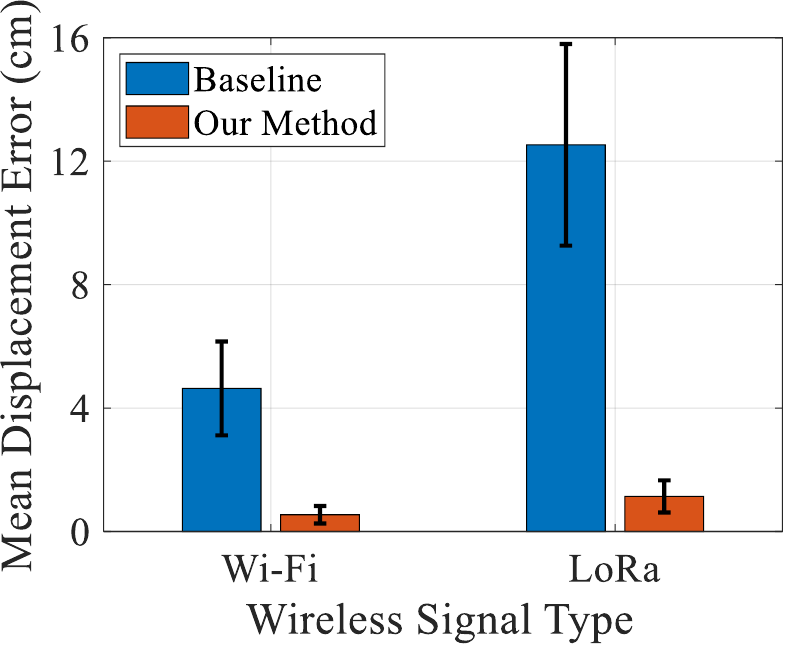}
        \vspace{-1mm}
        \caption{\small Gesture experiment results.}
        \vspace{-0mm}
        \label{fig:fig_7}
    \end{minipage}
\end{figure*}

\section{Experiment}
To evaluate our proposed framework, we conduct real-world experiments using both Wi-Fi and LoRa signals. 
Specifically, we evaluate the performance of our framework in estimating target displacement through a benchmark experiment and a gesture experiment.

\textbf{Wi-Fi Setup.} We use two mini PCs equipped with a commodity Intel 5300 wireless NIC to transmit and receive Wi-Fi signals. The transmitter is configured with a single antenna, while the receiver is equipped with two antennas. The Wi-Fi signals are centered at a frequency of 2.47~GHz, corresponding to a wavelength of 12.1~cm. We collect Wi-Fi Channel State Information (CSI) as channel measurements using the CSI Tool~\cite{halperin2011tool}, at a sampling rate of 1000~Hz.

\textbf{LoRa Setup.} We use a Semtech SX1276 module with an Arduino Uno, equipped with a single antenna, as the LoRa node. The LoRa gateway is implemented using a USRP B210 equipped with two antennas and GNU Radio. The LoRa node transmits signals at a center frequency of 915~MHz, corresponding to a wavelength of 32.8~cm. We dechirp the received LoRa signals to obtain channel measurements at a sampling rate of 1000 Hz.

\textbf{Baseline.}
We employ the state-of-the-art method proposed in~\cite{gao2022towards} as the baseline. Our framework follows the baseline method for extracting the rotation angle of the channel ratio, but the baseline does not incorporate our sub-wavelength accuracy recovery scheme.

\subsection{Benchmark Experiment}
In the benchmark experiment, we use a 30~cm $\times$ 30~cm metal plate as the target.
A Wi-Fi transmitter and receiver are deployed 1.2~m away from the metal plate, with a spacing of 4~m between them. 
The metal plate is mounted on a high-precision sliding track with a digital controller, enabling displacement control with an accuracy of 1~mm. 
We configure the metal plate to move along the perpendicular bisector of the transceiver pair at speeds of 1~cm/s, 2~cm/s, and 5~cm/s, with displacement distances of 2~cm, 5~cm, 10~cm, 15~cm, and 20~cm. 
Each experiment is repeated 20 times.

Figure~\ref{fig:fig_5} shows an example of displacement estimation results over time for a single target movement experiment.
We observe that when the target path length change is not an exact integer multiple of the wavelength, the baseline method exhibits periodic over- or under-estimation relative to the sliding track ground truth, with a maximum error of 6.3~cm.
In contrast, our method maintains displacement estimates consistent with the ground truth at the sub-wavelength scale, with a maximum error below 0.9~cm.

Figure~\ref{fig:fig_6} presents the Cumulative Distribution Functions~(CDFs) of displacement estimation errors across all benchmark experiments. It can be observed that our method reduces the median error from 2.7~cm to 0.3~cm, and the 90th-percentile error from 5.3~cm to 0.7~cm. These results demonstrate that our framework significantly advances the sensing accuracy limits at the sub-wavelength scale.

\subsection{Quantifying Gesture Displacement}
To evaluate the potential of our framework in sensing applications, we conduct gesture experiments using both Wi-Fi and LoRa devices. In these experiments, the transmitter and receiver are placed on a table in an office environment, separated by 2~m. Volunteers perform gestures from an initial position 1~m away from the transceivers. To obtain the ground truth, we pre-mark reference points on the table at displacements of 5~cm, 10~cm, 15~cm, and 20~cm, and instruct the volunteers to move their hands between these points. We recruit four volunteers~(two males and two females) to participate in the experiments, and each experiment is repeated 10 times.

Figure~\ref{fig:fig_7} presents the average displacement estimation errors of our method compared with the baseline. Notably, our method achieves nearly an order-of-magnitude improvement in accuracy for both Wi-Fi and LoRa sensing. Specifically, when sensing with Wi-Fi signals, the average errors of the baseline and our framework are 4.6~cm and 0.5~cm, respectively. When using LoRa signals, which have a longer wavelength, the average error of the baseline increases to 12.5~cm, while our method maintains a low average error of 1.1~cm. These results demonstrate that our framework significantly enhances the sub-wavelength accuracy of gesture displacement estimation, opening new opportunities such as fine-grained control of IoT devices via in-air gestures.
\section{Conclusion}
In this paper, we address the fundamental sub-wavelength accuracy limitation in bistatic wireless sensing based on the channel ratio.
Beyond prior qualitative understanding, we derive the first quantitative mapping between the distorted sensing feature of the channel ratio and the ideal sensing feature.
We reveal that this mapping is determined only by the amplitude of channel responses, which is unaffected by detrimental phase offsets in bistatic wireless systems.
Building on this foundation, we propose a robust framework to recover the ideal sensing feature from the distorted channel ratio.
Real-world experiments across Wi-Fi and LoRa demonstrate that our framework achieves nearly an order-of-magnitude accuracy improvement at the sub-wavelength scale.

\bibliographystyle{IEEEbib}
\bibliography{strings, refs}

\end{document}